\documentclass[10pt, conference, letterpaper]{IEEEtran}
\ifCLASSINFOpdf
\else
\fi
\usepackage{lipsum}
\usepackage{graphicx}
\usepackage{float}
\usepackage{cite}
\usepackage{epstopdf}
\usepackage{multirow}
\usepackage{amssymb}
\usepackage{amsmath}

\usepackage{mathtools, cases}
\usepackage{chngcntr}
\counterwithin*{equation}{section}
\usepackage{txfonts}
\usepackage{amsmath,etoolbox}
\usepackage{algorithmic}
\usepackage{amssymb}
\usepackage[ruled]{algorithm2e}
\usepackage{url}
\usepackage[mathscr]{euscript}

\usepackage{caption}
\usepackage{subcaption}
\usepackage[font=small]{caption}
\usepackage{cleveref}
\usepackage{color}

\begin{document}
%
\title{WiLiTV: A Low-Cost Wireless Framework for Live TV Services}
\author{$\text{Rajeev Kumar}^{\dagger}, \text{Robert S Margolies}^{*}, \text{Rittwik Jana}^{*}, \text{Yong Liu}^{\dagger} \text{and} \ \text{Shivendra Panwar}^{\dagger}$\\
${\dagger}$ Department of Computer and Electrical Engineering, New York University, NY 11201, USA \\
${*}$ AT\&T Labs Research, 1 AT\&T Way, Bedminster, NJ 07921, USA \\ 
}
\maketitle

\begin{abstract}
With the evolution of HDTV and Ultra HDTV, the bandwidth requirement for IP-based TV content is rapidly increasing. Consumers demand uninterrupted service with a high Quality of Experience. Service providers are constantly trying to differentiate themselves by innovating new ways of distributing content more efficiently with lower cost and higher penetration. In this work, we propose a cost-efficient wireless framework (WiLiTV) for delivering live TV services, consisting of a mix of wireless access technologies (e.g. Satellite, WiFi and LTE overlay links). In the proposed architecture, live TV content is injected into the network at a few residential locations using satellite dishes. The content is then further distributed to other homes using a house-to-house WiFi network or via an overlay LTE network. Our problem is to construct an optimal TV distribution network with the minimum number of satellite injection points, while preserving the highest QoE, for different neighborhood densities. We evaluate the framework using realistic time-varying demand patterns and a diverse set of home location data. Our study demonstrates that the architecture requires 75--90\% fewer satellite injection points, compared to traditional architectures. Furthermore, we show that most cost savings can be obtained using simple and practical relay routing solutions.
\end{abstract}
\IEEEpeerreviewmaketitle

\section{Introduction}
\label{sec:intro}
\begin{figure}[t]
         \centering
         \begin{subfigure}[b]{0.45\textwidth}
                 \includegraphics[width=\textwidth]{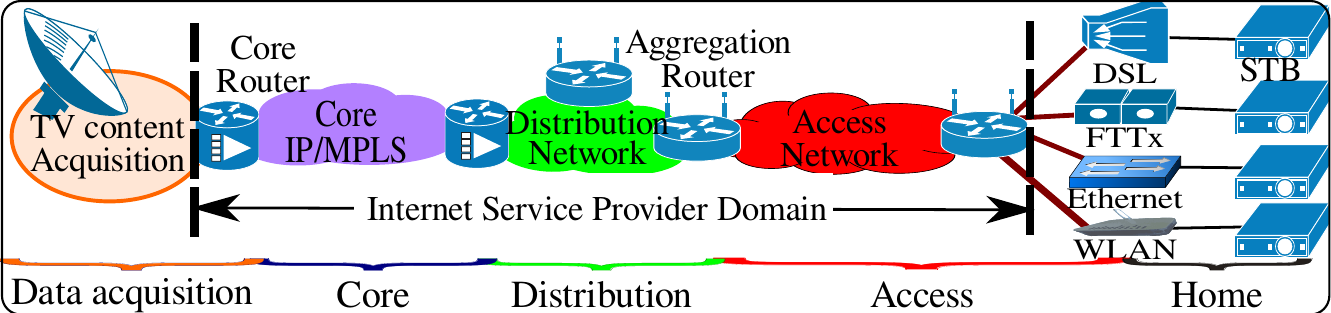}
                 \caption{Traditional IPTV architecture}
                 \label{fig:WiLiTV}
         \end{subfigure}\\
         \begin{subfigure}[b]{0.45\textwidth}
                 \includegraphics[width=\textwidth]{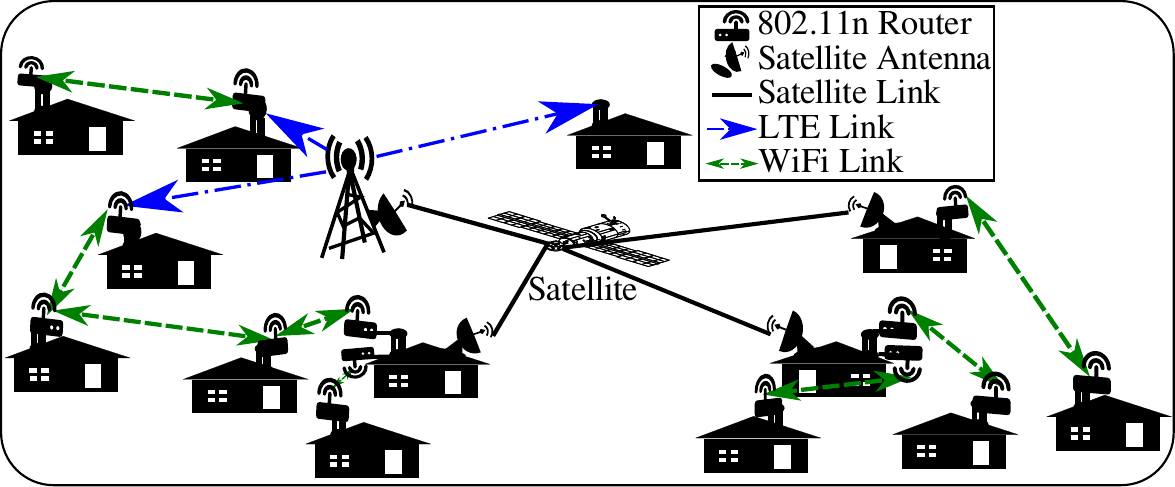}
                 \caption{\emph{Our proposed WiLiTV architecture:-} content is first delivered to a community through a few selected houses and LTE BSs with satellite antennas, and is then distributed to houses using WiFi and LTE relays.}
                 \label{fig:topology}
         \end{subfigure}%
         \caption{Comparison of IPTV and our proposed WiLiTV architectures.}
  \label{fig:TV}
\vspace{-4mm}
\end{figure}

Today, the vast majority of households receive TV content via cable/fiber, IP network, or satellite. As illustrated in Fig.~\ref{fig:TV}(\subref{fig:WiLiTV}), Internet Protocol TV (IPTV) streams live TV content from a few regional hub offices to set-top boxes over either a dedicated private network or over-the-top via the core IP network~\cite{rf1}.
To satisfy Quality of Service (QoS) requirements, IPTV must be provisioned with a sufficiently high bandwidth in the distribution network~\cite{rf34}.

However, with the evolution of HDTV, 4K content, and the prevalence of thousands of channels, the need for bandwidth is ever increasing~\cite{rf35}. Even with advanced video compression techniques, each Standard and High Definition TV (SDTV, HDTV) channel requires 2 and 9 Mbps, respectively. Thus, the current infrastructure will soon be stressed with this escalating demand.
 
 
One solution to the growing demand is to deploy more cables/fibers. However, deployment of wired infrastructure is costly, especially in rural areas. The per-house fiber-laying cost can go up to \$19,000 if the number of households per mile is less than five~\cite{rf40}.
Another possible solution is to scale up the capacity of the IP core to handle the increased traffic; however, this will require additional routing equipment, thus resulting in greater infrastructure cost~\cite{rf42} and greater energy consumption~\cite{rf3}. Satellite TV providers avoid the wired infrastructure cost by broadcasting live TV content to \emph{every} subscriber/household equipped with a satellite dish antenna. However, satellite providers still incur a high initial cost to install a dish antenna for each new customer household. In New York, for example, installing a single satellite dish costs approximately \$1,000~\cite{rf39}.
 
Service providers constantly strive to differentiate themselves by innovating new ways of distributing content more efficiently with lower cost and higher penetration.
Therefore, in this paper, we propose a low-cost, Wireless Live TV (WiLiTV) architecture that leverages a range of access technologies (Satellite, WiFi and LTE) to provide high quality live TV services. As shown in Fig.~\ref{fig:TV}(\subref{fig:topology}), the WiLiTV architecture strategically equips a few households and/or LTE Base Stations (BSs) with satellite dish antennas and relays TV content to other residential homes using WiFi and/or cellular networks. Our proposed architecture offloads TV content from the traditional core IP network or a dedicated wired IPTV infrastructure to long-haul satellite links and local high speed wireless links among households, leveraging the recent advances in wireless technologies, such as Massive MIMO and Millimeter Wave. With this novel architecture, our design goal is to satisfy the live TV demands of all households at the lowest possible infrastructure cost. It consists of two sub-problems:
\begin{itemize}
\item {\it Source Provisioning:} which households are chosen to install satellite antennas and download all live TV channels?
\item {\it Relay Routing:} how should live TV channels requested by each household be relayed from the sources?
\end{itemize}
These two sub-problems are tightly coupled: source provisioning determines the potential sources that a household can download content from; and if some households cannot find a relay routing solution to get their desired channels, new sources have to be added to the distribution network. There is a fundamental trade-off between the complexity of relay routing and the cost saving in source provisioning. The current satellite TV providers, such as Dish Network~\cite{rf41}, is at one end of the trade-off spectrum, where each household installs a satellite antenna and no relay routing is needed at all. At the other end, one can minimize the number of satellite antennas to be installed by maximally utilizing any possible wireless relay routing among households to satisfy their live TV demands. Any practical solution has to find the sweet point and strike the right balance between complexity and cost-saving. To systematically evaluate the impact of various relay routing factors on cost saving, we formulate a series of joint provisioning-routing optimization problems  to find the lowest costs under different routing constraints, including relay hop count limit, splittable or non-splittable flows, LTE availability, and dynamic or static solutions. The optimization models developed are used to numerically investigate the routing complexity and cost saving tradeoff through case studies with real household topology and user demand data.
The key contributions of this paper are as follows,
\begin{enumerate}
\item We present the WiLiTV architecture to provide high quality live TV services to users via a mix of wireless access technologies (i.e., satellites, WiFi relayed communication, and LTE overlay network).
\item We formulate a series of novel joint optimization problems to systematically evaluate the trade-off between the cost saving in satellite antenna provisioning and the complexity in relay routing, considering various practical provisioning and routing constraints.
\item The formulated optimization problems can be solved using binary programming, or mixed-integer programming for small and medium networks. For large networks, we develop greedy heuristic algorithms to obtain close-to-optimal solutions.
\item We evaluate the proposed WiLiTV architecture using real household topology and user demand data from a major live TV service provider. Our results demonstrate that WiLiTV requires 75\% to 90\% fewer satellite injection points, compared to traditional architectures. Most of the cost savings can be achieved with simple and practical relay routing solutions.
\end{enumerate}
This paper is organized as follows. In Section~\ref{sec:RelatedWork} we discuss the related work. The system model and assumptions made are described in Section \ref{sec:system-model}. The formulation of joint optimization problems are presented in Section \ref{sec:optimization}. Section~\ref{sec:Solution} and Section~\ref{sec:results} contain the associated solution techniques and the numerical results, respectively. Section \ref{sec:conclusion} concludes the paper.

\section{Related Work}
\label{sec:RelatedWork}
An IPTV architecture can be divided into five main parts, (i) a data acquisition network, (ii) core backbone network containing super hub offices (SHO), (iii) a regional distribution network containing video hub offices (VHO), (iv) access network containing DSLAMs, and (v) customer home network containing residential gateways and set-top-boxes~\cite{rf8,rf10}. To decrease bandwidth requirements in the core backbone network and for fast TV channel switching, multicast channels and groups are typically used. However, maintaining multicast puts an extra burden on the network, especially for building and pruning multicast groups. Furthermore, IP multicasting is a good candidate for popularly viewed channels. On the other hand, it is costly to maintain multicast groups for distributing less popular TV channels~\cite{rf15}. Peer-to-peer is another technology that has been investigated for distributing live TV~\cite{rf10}. However, there are challenges associated with P2P for accommodating fast TV channel switching and TV channel recovery, especially when streaming peers leave the system abruptly. This can result in interruption while viewing live TV and eventually a poor QoE. We are thus motivated to find a suitable architecture which pushes TV content to the end users using a diverse set of access links simultaneously. This diversity is useful to provide resiliency and bandwidth aggregation to satisfy the cumulative demand from the end users. Our proposed WiLiTV architecture is able to push TV content to the households, facilitate fast channel switching while also providing a cost-effective solution for distribution of less popular TV channels.
 
Several studies present the challenges of and solutions for providing live TV services with QoS guarantees using wireless technologies (WiFi, WiMax, etc.)~\cite{rf29,rf11,rf30,rf31,rf32,rf33}. Note that in all these studies, TV content is delivered to the access network through SHO and VHO, which results in a large bandwidth requirement and energy consumption in the distribution network. By contrast, our solution does not need the backbone network and offloads TV traffic to local wireless networks.  
\vspace{-4mm}       
\section{System Model and Assumptions}
\label{sec:system-model}
As illustrated in Fig.~\ref{fig:TV}(\subref{fig:topology}), our wireless distribution network for live TV consists of three types of nodes:
\begin{enumerate}
\item A small number of households equipped with satellite antennas act as the injection points for live TV content. They also have WiFi access points for relaying content to WiFi-only households.

\item LTE BSs equipped with satellite antennas act as additional live TV injection points, and can deliver content to LTE-enabled households over unused LTE bands.

\item Regular households are equipped with WiFi access points and LTE receivers. A regular household receives TV content from WiFi or LTE. It can also relay the received TV content to other regular households.
\end{enumerate}

\noindent As a result, a household can receive TV content by the following methods: (i) directly from satellite antenna, (ii) through WiFi relay, (iii) through  LTE relay, and (iv) through both LTE and WiFi relays. Fig.~\ref{fig:TV}(\subref{fig:topology}) illustrates the TV reception and relay methods at each node. Moreover, content can be relayed using either {\it all-or-nothing flows} or {\it fractional flows}. In all-or-nothing flows, a household receives all content from a single source/relay node; using fractional flows, a household receives content simultaneously from multiple sources/relays. TV traffic demand at household $i$ is denoted by $\delta_i$ (in Mbps). The demand can also be expressed as $\psi_i*b$, where $\psi_i$ is the numbers of channels being demanded at household $i$  and $b$ is the capacity required per channel in Mbps.
\subsection{Relay using WiFi}
\label{subsec:WiFi}
The WiFi relay network is modeled as an undirected graph $\mathcal{G}=(\mathcal{V},\mathcal{E})$, where $\mathcal{V}$ is the set of households and $\mathcal{E}$ is the set of WiFi links between households. WiFi transmissions between neighboring households operate on orthogonal channels, and are highly directional by making use of beam-forming techniques. Point-to-point connections among households avoid wasting airtime in collision avoidance. Furthermore, the households are bounded by a degree of connectivity represented by $\rho$, i.e., a household has a maximum of $\rho$ point-to-point links with neighboring households. We assume all WiFi transmitters have the same transmit power $P$, and path losses ($PL$) between two households are the same along both directions ($PL_{ij}=PL_{ji}$, between household $i$ and $j$)  \cite{rf18}. A WiFi link exists from household $i$ to $j$ if $j$ lies within the communication range of $i$; specifically, if the received signal strength on $j$ is greater than the receiver sensitivity \cite{rf19},
\begin{equation}
 P - PL_{ij}\geq\xi; \forall i,j \in\mathcal{V},
\label{eqn:radio sensitivity}
\end{equation}
where $\xi$ is the WiFi receiver sensitivity, and it is assumed to be identical for all WiFi receivers. According to \cite{rf20}, the pathloss on a WiFi link can be calculated as
\begin{equation}
\label{eqn:pathLossWiFi}
PL(d)=
\begin{cases} L_{FS}(d)+\mathit{SF}; \text{ if } d< d_{BP}, \\
L_{FS}(d_{BP})+35log(\frac{d}{d_{BP}})+\mathit{SF};\text{ if } d\geq d_{BP},
\end{cases}
\end{equation}
where $d$ is the distance between the transmitter and receiver, $L_{FS}(d)$ is the free space pathloss in dB, $d_{BP}$ is the breakpoint distance and $\mathit{SF}$ is shadow fading in dB. The free space pathloss is defined as
\begin{equation}
 \label{eqn:FS}
 L_{FS}=20log(d)+20log(f)-147.5,
\end{equation}
where $f$ is the carrier frequency. From the transmit power and pathloss computed with equations \ref{eqn:pathLossWiFi}-\ref{eqn:FS}, the received signal strength at $j$ can be calculated. We use the tables in \cite{rf20,rf21} to map the received signal strength to the corresponding modulation and coding scheme and the achievable capacity of WiFi links. Since all transmitters have the same transmit power, and pathloss is  symmetric, we have  $C_{ij}=C_{ji} \ \text{for} \ i\neq j$.
\vspace{-5mm}
\subsection{Relay using WiFi and LTE}
\label{subsec:LTE}
LTE BSs can be additional injection points of live TV content, subject to the availability of LTE bandwidth at the BS. Let $\mathcal{L}$ indicate the set of LTE BSs having significant spare LTE resources. The network topology is augmented as
$\mathcal{G}^{'}=(\mathcal{V}^{'},\mathcal{E}^{'})$, with $\mathcal{V}^{'}=(\mathcal{V} \cup \mathcal{L})$ and $\mathcal{E}^{'}$ consisting of all WiFi and LTE links. LTE BSs can only be a source node. Thus, all LTE links in the topology are unidirectional from a LTE BS to households. A LTE BS uses a single channel for transmission in its coverage area. Thus, resources must be shared between households receiving TV content from the same LTE BS. We use TDMA for resource sharing. Let $0 \leq \lambda_{ij}\leq 1$ be the time share of the link from LTE BS $i \in \mathcal{L}$ to household $j\in \mathcal{V}$, $\sum_j \lambda_{ij} \le 1$, $\forall i \in \mathcal{L}$. To characterize LTE links, the pathloss from LTE BS to households is calculated using \cite{rf22},
\begin{align}
 \label{eqn:pathLossLTE1}
 PL_{ij}^{LOS}=103.8+20.9log(d)
\end{align}
 where (\ref{eqn:pathLossLTE1}) represents pathloss from an LTE BS to a household for the line-of-sight link. Using the maximum allowed transmit power of LTE BSs and the pathloss model, we evaluate the LTE capacity as~\cite{rf23},
\begin{equation}
 \label{eqn:Cap}
 C_{ij}^{LTE}=\beta\mathcal{W}log_2(1+\gamma {SNR}),
\end{equation}
 where $\beta$ is the fraction of bandwidth used for data transmission while the rest is used for control signaling. Typically, $\beta$ ranges between $0.5-0.8$. Similarly, $\gamma$ is the fraction of received signal to noise ratio that contributes to broadband speed. Typically, $\gamma$ lies between $0.5$ to $0.6$.

For easy reference, all the notation is presented in Table \ref{notations}.
\begin{table}[!ht]
\renewcommand{\arraystretch}{1.3}
\caption{Notation}
\label{notations}
\centering
\begin{tabular}{|p{4em}|p{24em}|}
\hline
\textbf{Parameter}      & \textbf{Description} \\ \hline
$\mathcal{V},\mathcal{L},\mathcal{V}^{'}$ 	& Set of households, LTE BSs and both, respectively \\ \hline
$\mathcal{E},\mathcal{E}^{'}$   & Set of WiFi links, set of WiFi and LTE links \\ \hline
$\mathcal{S},\mathcal{R},\mathcal{T}$  	& Set of Source, Relay and Terminal nodes, respectively \\ \hline
$\delta_i$              	& Demand at household $i$  \\ \hline
$h$                     	& Maximum allowed hops in the topology  \\ \hline
$\rho$                  	& Maximum degree of connectivity at source and relay nodes  \\ \hline
$C_{ij}$                	& Capacity of link from node $i$ to $j$  \\ \hline
$u_{ij}$                	& Binary variable indicating if link from node $i$ to $j$ is selected for content distribution \\ \hline
$X_i$                   	& Binary variable indicating if household $i$ has a satellite antenna \\ \hline
$Y_i$                   	& Binary variable indicating if household $i$ relays video to neighboring households \\ \hline
$l_{si}$                	& Binary variable indicating if node $i$ downloads video directly from virtual source $s$ (fractional flow) \\ \hline
$f_{ij}$                	& Video traffic on link from $i$ to $j$ (fractional flow) in Mbps\\ \hline
$\lambda_{ij}$          	& Time share of node $j$ from LTE BS $i$\\ \hline
$\Delta_i(t)$           	& Binary variable indicating if household $i$ requires TV services at time instance $t$\\ \hline
$\tau_i$           		& Available resources at the LTE BS $i$\\ \hline
\end{tabular}
\vspace{-4mm}
\end{table}
\section{Joint Optimization of Satellite Antenna Placement and Relay Routing}
\label{sec:optimization}
In this section, we develop optimization models to systematically evaluate the design trade-offs in WiLiTV. We consider the following routing complexity factors.
\begin{enumerate}
\item {\it Relay Hop Count:} Ideally, each connected island of households only needs one source, and TV content can be relayed to all households using an arbitrary number of hops. However, live TV services have stringent QoS requirements on delay, bandwidth and reliability. It is well known that multi-hop wireless relays can lead to long delay, low end-to-end throughput and poor reliability \cite{rf27}. In this paper, we limit relay routing to be at most two hops. We will compare the cost saving with one-hop and two-hop relay routing.

\item {\it Splittable Flows:} As discussed in Section \ref{sec:system-model}, with fractional flows, one household can download content from multiple relay paths from multiple sources. This can potentially increase the wireless link utilization and coverage of each source, leading to higher cost saving. As with any multi-path routing, splittable flows have to deal with delay disparity on different paths, and data transmission reliability decreases as more links and nodes are employed. We will compare the efficiency of relay routing with and without splittable flows.

\item {\it LTE Availability:} A LTE BS can cover a wider range than a WiFi transmitter. But LTE bandwidth resources are expensive. We will evaluate the coverage gain added by LTE BS to justify its bandwidth cost.

\item {\it Dynamic vs. Static Solution:} User TV demands naturally vary over time. To maximally reduce cost, one should design dynamic source provisioning and relay routing solutions to match the changing user demand. However, it is not practical to change satellite antenna locations on an hourly or daily basis, and reconfiguring relay routing may cause short-term service disruption. Static solutions are easier to implement. We will formally study the performance gap between dynamic and static solutions.
\end{enumerate}
To systematically evaluate the impact of various source provisioning and relay routing strategies on cost saving, we formulate a series of joint provisioning-routing optimization problems to find the lowest costs under different routing constraints in this section. 
%
%
\vspace{-2mm}
\subsection{Fixed Demand with WiFi}
\label{subsec:optA}
We start with the simple scenario where user demands are fixed and only WiFi relays are available. We use the graph $\mathcal{G}$ having only WiFi transmitters and receivers. We first formulate the optimization problems for non-splittable flow routing with one-hop and two-hop relays respectively. We then generalize it to splittable flow routing with limited hop count.
\subsubsection{One-hop and Non-splittable Relay Routing}
\label{subsubsec:op1}
For this scenario, we assume that a household is at most one hop apart from its corresponding source node. Let $X_i \in \{0,1\}$, $\forall i \in \mathcal{V}$  be the binary variable indicating whether a node is equipped with satellite antenna. Similarly, $u_{ij} \in \{0,1\}$, $\forall \langle i, j \rangle \in \mathcal{E}$ be the binary variable indicating if a link from node $i$ to node $j$ carries node $j$'s TV demands. Using these binary variables, we can formulate a binary programming problem as follows:
\begin{alignat}{2}
\label{eqn:op1}
\MoveEqLeft[2]\underset{\{X_i, u_{ij}\}}{\text{\bf Minimize:}} \sum_{i \in \mathcal{V}} X_i \\
\label{eqn:opt1_1}
\MoveEqLeft \text{\bf Subject to:}\sum_{j: \langle i, j \rangle \in \mathcal{E} }u_{ij} \leq \rho X_i, \ \forall i \in \mathcal{V}; \\
\label{eqn:opt1_2}
\MoveEqLeft \qquad\qquad \ \sum_{i: \langle i, j \rangle \in \mathcal{E}  }u_{ij} = (1-X_j), \ \forall j \in \mathcal{V}; \\
\label{eqn:opt1_3}
\MoveEqLeft \qquad\qquad \ \sum_{i: \langle i, j \rangle \in \mathcal{E} } u_{ij} C_{ij} \geq \delta_j (1-X_j), \ \forall j \in \mathcal{V}.
\end{alignat}
The objective (\ref{eqn:op1}) is to minimize the number of satellite antennas. Constraint (\ref{eqn:opt1_1}) dictates that if node $i$ is selected as a source node ($X_i=1$), the number of its receivers is bounded by the degree of connectivity $\rho$; otherwise ($X_i=0$), node $i$ cannot have any out-going video traffic.
Constraint (\ref{eqn:opt1_2}) reflects the fact that a non-source node downloads video content from exactly one incoming link, and a source node does not have any incoming video traffic. Constraint (\ref{eqn:opt1_3}) states that, at a non-source household, the aggregate bandwidth of all incoming links must be greater than its total TV demand.
\subsubsection{Two-hop and Non-splittable Relay Routing}
\label{subsubsec:op2}
\begin{figure}[t]
\centering
\includegraphics[width=0.45\textwidth]{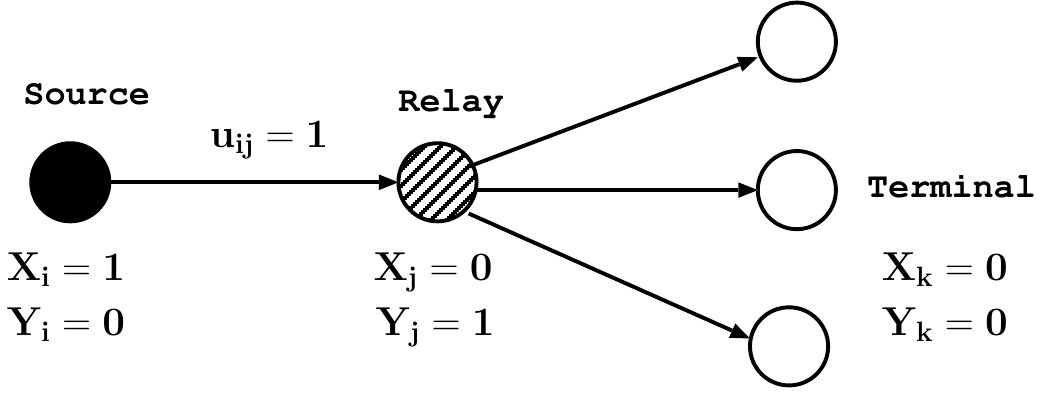}
\caption{Two-hop Relay Routing: node $i$ is a source node ($X_i=1$), node $j$ is a relay ($Y_j=1$), downloading traffic from $i$ ($u_{ij}=1$), relaying video to other terminal nodes ($X_k=Y_k=0$).}
\label{fig:flow}
\vspace{-5mm}
\end{figure}

Now we relax the maximum relay hop count to two. Thus, some non-source households may relay video traffic for other non-source households. There are three types of households in the network:  source nodes with satellite antennas, non-source nodes relaying video for other nodes (called relay nodes), and non-source nodes without any relaying traffic (called terminal nodes). Using $X_i$, introduced in the previous formulation, all non-source nodes have $X_i=0$. We further introduce another binary variable $Y_i \in \{0,1\}$, $i \in \mathcal{V}$ to indicate whether a node relays other nodes' traffic. Then for a relay node we have $X_i=0$ and $Y_i=1$, and for a terminal node, we have $X_i=0$ and $Y_i=0$. Fig.~\ref{fig:flow} illustrates the three types of nodes in the two-hop relay, and how terminal nodes download video content from the source through a common relay node. The joint optimization problem with two-hop relay can be formulated as a new binary programming problem:
\begin{alignat}{2}
\label{eqn:op2}
\MoveEqLeft[2]\underset{\{X_i, Y_i, u_{ij}\}}{\text{\bf Minimize:}} \sum_{i \in \mathcal{V}} X_i \\
\MoveEqLeft \text{\bf Subject to:}\notag\\
\label{eqn:opt2_1}
\MoveEqLeft \sum_{j: \langle i, j \rangle \in \mathcal{E} } u_{ij} \leq \rho (X_i+Y_i), \ \forall i \in \mathcal{V}; \\
\label{eqn:opt2_2}
\MoveEqLeft \sum_{i: \langle i, j \rangle \in \mathcal{E}  } u_{ij} = (1-X_j), \ \forall j \in \mathcal{V}; \\
\label{eqn:opt2_3}
\MoveEqLeft 0 \leq X_i+Y_i \leq 1, \ \forall i \in \mathcal{V}; \\
\label{eqn:opt2_4}
\MoveEqLeft Y_j \leq 2-Y_i-u_{ij}, \ \forall i, j \in \mathcal{V}; \\
\label{eqn:opt2_5}
\MoveEqLeft u_{ij} C_{ij} \geq \delta_j u_{ij}+\sum_{k: k\neq i,j} \delta_k u_{jk} - \Theta(1-X_i-Y_i), \ \forall \langle i, j \rangle \in \mathcal{E}.
\end{alignat}
Constraint (\ref{eqn:opt2_1}) bounds the maximum degree of connectivity at source and relay nodes (both have $X_i+Y_i=1$), and terminal nodes cannot have outgoing video traffic ($X_i+Y_i=0$). According to constraint (\ref{eqn:opt2_2}), all non-source nodes download their video from exactly one incoming link. Constraint (\ref{eqn:opt2_3}) states that a node in the distribution network can only assume one role out of source, relay or terminal node. Constraint (\ref{eqn:opt2_4}) enforces that a relay node does not receive traffic from another relay node. This is because if node $j$ receives video from a relay node $i$, then $Y_i=1$ and $u_{ij}=1$. Then to make (\ref{eqn:opt2_4}) hold, we must have $Y_j=0$, i.e, $j$ cannot be a relay node anymore. On the other hand, if $i$ is a source node, $Y_i=0$, even if $u_{ij}=1$, we can still have $Y_j=1$ (i.e., $j$ can still relay video to other nodes). The last constraint guarantees each outgoing wireless link from a source or relay has enough bandwidth to carry video traffic assigned to it.
The first term on the righthand side is the video traffic from the source/relay node to its direct receiver. The second term is non-zero only if $i$ is a source and $j$ is a relay; it represents the traffic of all households downloading video from $i$ through relay $j$. The last term is zero if $i$ is a source or relay, and if $i$ is a terminal node,  $\Theta$ is a large number so that the inequality automatically holds.
\subsubsection{Splittable Relay Routing with Average Hop-count Limit}
\begin{figure}[t]
\centering
\includegraphics[width=0.45\textwidth]{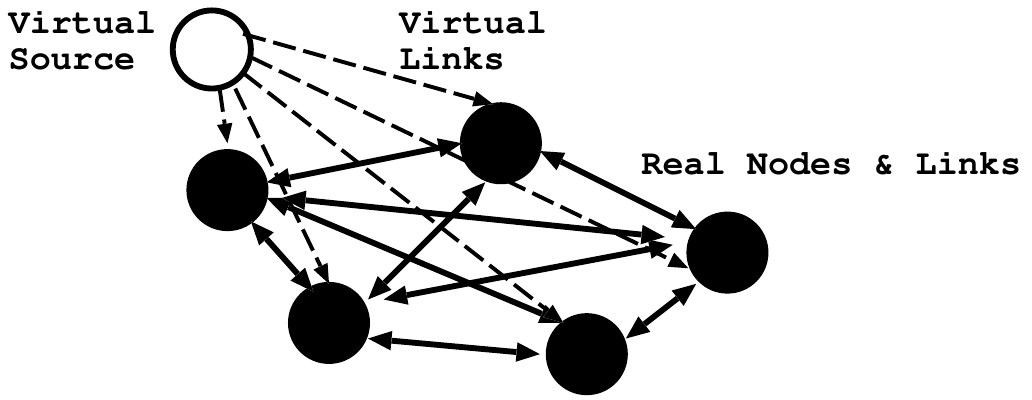}
\caption{Virtual Network Topology for Splittable Relay Routing.}
\label{fig:virtual}
\vspace{-5mm}
\end{figure}

In the last two optimization formulations, we considered all-or-nothing flows and each household downloads all its video demands from one source/relay through one wireless link. To further improve the flexibility and efficiency of relay routing, a household can receive video from multiple sources and/or relays simultaneously through multiple relay paths. We develop a variation of the well-known multi-commodity flow problem to cover this case.  As illustrated in Fig.~\ref{fig:virtual}, we first augment the distribution network with a virtual source node (indicated by $s$), that connects to all the nodes in the topology through virtual links with very high capacities. All video demands are served from the virtual source. If a household installs a satellite antenna, it is equivalent to saying that we activated its virtual link from the virtual source for direct video downloading. The objective of minimizing the number of satellite dishes is equivalent to minimizing the number of activated virtual links. We define a binary variable $l_{si}$ indicating whether the virtual link from the virtual source $s$ to node $i$ is activated. We further define $f_{ij}$ as the video traffic volume on link $\langle i, j \rangle$. The optimization with splittable relay routing can be formulated as the following mixed-integer programming problem.
\begin{alignat}{2}
\label{eqn:op3}
\MoveEqLeft[2]\underset{\{l_{si}\},\{f_{ij}\}}{\text{\bf Minimize:}} \sum_{i \in \mathcal{V}} l_{si} \\
\MoveEqLeft \text{\bf Subject to:} \notag\\
\label{eqn:op3-1}
\MoveEqLeft f_{si} + \sum_{j: \langle j, i \rangle \in \mathcal{E} } f_{ji}
= \delta_i + \sum_{k: \langle i, k \rangle \in \mathcal{E} } f_{ik}, \ \forall i \in \mathcal{V}; \\
\label{eqn:op3-2}
\MoveEqLeft \sum_{i \in \mathcal {V}} f_{si} = \sum_{i \in \mathcal {V}} \delta_i; \\
\label{eqn:op3-31}
\MoveEqLeft f_{ij} \leq C_{ij}, \ \forall \langle i, j \rangle \in \mathcal{E};\\
\label{eqn:op3-3}
\MoveEqLeft f_{si} \leq l_{si} C_{si}, \ \forall i \in \mathcal{V};\\
\label{eqn:op3-4}
\MoveEqLeft \sum_{\langle i, j \rangle \in \mathcal{E}} f_{ij} \leq h \sum_{i \in \mathcal {V}} \delta_i, \ \forall i \in \mathcal{V}.
\end{alignat}
Constraint (\ref{eqn:op3-1}) is the {\it flow-conservation} law at node $i$, i.e., the total incoming traffic at node $i$ (left-hand side) equals the sum of the demand of node $i$ and the total outgoing traffic (right-hand side).  (\ref{eqn:op3-2})  implies that all the video downloading traffic in the virtual graph originates from the virtual source. (\ref{eqn:op3-31}) guarantees traffic on each relay link is bounded by its capacity, and (\ref{eqn:op3-3})  makes sure that a virtual link can carry video traffic only if it is activated. Finally, the left-hand side of (\ref{eqn:op3-4}) is the total video traffic on all relay links, i.e., the sum of the traffic generated by all households on all links. For each household, the total traffic it generates on all links equals its total video demand multiplied by its average relay hop count. (\ref{eqn:op3-4}) effectively limits the average relay hop count of all households to a constant $h$.
\label{subsubsec:op3}
\subsection{Fixed Demand with WiFi and LTE}
\label{subsec:optB}
As discussed in Section~\ref{subsec:LTE}, an LTE BS can be a potential injection point of live TV content.  We now discuss the problem formulation with LTE BSs by extending the formulations in the previous section.
\subsubsection{One-hop and Non-splittable Relay}
\label{subsubsec:op4}
As modeled in Section~\ref{subsec:LTE}, we extend the distribution network from $\mathcal{G}$ to $\mathcal{G}^{'}=(\mathcal{V}^{'},\mathcal{E}^{'})$ by including LTE BSs and LTE links from LTE BSs to their covered households. The WiFi optimization problem defined in  (\ref{eqn:op1}) to (\ref{eqn:opt1_3})
 can be extended to cover the LTE case. Each LTE BS can be a potential injection point, we extend $X_i$ defined on $\mathcal{G}$ to $\mathcal{G}^{'}$, $X_i=1$, $\forall i \in \mathcal L$, if and only if we install satellite antenna on LTE BS $i$. Then the optimization objective is to minimize the number of satellite antennas among LTE BSs and households, i.e.,   $ \min \sum_{i \in \mathcal{V'}} X_i$. Constraints for households defined in (\ref{eqn:opt1_1}), (\ref{eqn:opt1_2}) and (\ref{eqn:opt1_3}) still hold. We introduce additional constraints for LTE BS:
 \begin{align}
\label{eqn:adc1}
\MoveEqLeft \sum_{j \in \mathcal V} \lambda_{ij} \leq X_i\tau_i, \ \forall i \in \mathcal{L}; \\
 \label{eqn:adc2}
\MoveEqLeft \lambda_{ij} C_{ij} \geq \delta_j u_{ij}, \ \forall i \in \mathcal{L}, \forall j \in \mathcal{V}.
\end{align}
The constraint in (\ref{eqn:adc1}) states that if BS $i$ does not have a satellite antenna, households cannot download video from it; if it does, then the total time shares of all covered household is bounded by available resources at the LTE BS. (\ref{eqn:adc2}) implies that the allocated bandwidth from BS $i$ to household $j$ is greater than the demand of $j$.
\subsubsection{Two-hop and Non-splittable Relay}
\label{subsubsec:op5}
In two-hop relay, LTE BSs can only be potential sources, so we can extend the optimization problem defined in (\ref{eqn:op2}) through (\ref{eqn:opt2_5}) by updating the objective function to $ \min \sum_{i \in \mathcal{V'}} X_i$, and adding constraints (\ref{eqn:adc1}) and a new LTE capacity constraint updated for two-hop relay:
\begin{equation*}
 \label{eqn:CapConsLTE}
\lambda_{ij} C_{ij} \geq \delta_j u_{ij}+\sum_{k,k\neq i,j} \delta_k u_{jk} - \Theta(1-X_i), \ \forall i \in \mathcal{L}, \forall j,k \in \mathcal{V}. \\
\end{equation*}
Similar to (\ref{eqn:opt2_5}), this constraint ensures that the link from BS $i$ to household $j$ carries video demands of household $j$ and all other households using $j$ as a relay. 
\subsubsection{Splittable Relay Routing}
\label{subsubsec:op6}
The inclusion of an LTE BS to the splittable relay routing formulation defined in (\ref{eqn:op3}) through (\ref{eqn:op3-4}) is straightforward by extending the objective function and all constraints to work on nodes and links in $\mathcal{G}^{'}=(\mathcal{V}^{'},\mathcal{E}^{'})$. The only change is that for a LTE link, the link capacity constraint (\ref{eqn:op3-31}) becomes:
\[f_{ij} \leq \lambda_{ij}C_{ij}, \quad \forall i \in \mathcal L, j \in \mathcal V,\]
reflecting that a LTE link is only active for a fraction of time.
\vspace{-5mm}
\subsection{Time Varying Demand}
\label{subsec:optC}
So far, the formulations assume user TV demands $\{\delta_i, i \in \mathcal V \}$ are fixed. In reality, user demands naturally vary over time. Let $t=1,\cdots,T$ be the typical time periods, and $\{\delta_i(t), i \in \mathcal V \}$ be the user demands at time period $t$. One approach is to design the distribution network to handle each user's maximum demand over all time periods, that is to let $\delta_i^0 \triangleq \max_{t=1,\cdots,T} \delta_i(t)$ and plug in the time-independent demands $\{\delta_i^0, i \in \mathcal V \}$ to the static formulations in the previous sections to obtain static provisioning and relay routing solutions. This over-provisioning might waste too much resources. In this section, we evaluate different ways to cope with time-varying user demands. Specifically, we consider the following cases: 1) dynamic provisioning of satellite antennas and dynamic relay routing; 2) static provisioning of satellite antennas and dynamic relay routing, and 3) static provisioning of satellite antennas and static relay routing. Satellite antenna installation cannot be easily adjusted on an hourly or daily basis. The first solution is not really practical. However, it gives us the lower bound on the required number of satellite antennas to meet time-varying user demands. The third solution may require more satellite antennas than the previous two. However, it is simpler to implement in practice. The second solution is practical and economical, since WiFi/LTE links and relay routing can be conveniently reconfigured using Software Defined Radio and/or Software Defined Networks.
\subsubsection{Dynamic Provisioning of Satellite Antennas and Dynamic Relay Routing}
\label{subsubsec:op7}
In a dynamic formulation, all the design variables $\{X_i, Y_i, u_{ij}, f_{ij}, l_{si},\lambda_{ij}\}$ in the static formulations should be converted to $\{X_i(t), Y_i(t), u_{ij}(t), f_{ij}(t), l_{si}(t), \lambda_{ij}(t)\}$. Other than the time-dependent demands $\{\delta_i(t), i \in \mathcal V\}$, we introduce another binary variable  $\Delta_i(t)$ such that if household $i$ has TV traffic demand at time $t$, then $\Delta_i(t)=1$, otherwise 0.
The one-hop and non-splittable relay routing problem defined in (\ref{eqn:op1}) through (\ref{eqn:opt1_3}) can be formulated for each time period $t$ as:
\begin{alignat}{2}
\label{eqn:op5}
\MoveEqLeft[2]\underset{\{X_i(t), u_{ij}(t)\}}{\text{\bf Minimize:}} \sum_{i \in \mathcal{V}} X_i(t) \\
\label{eqn:opt5_1}
\MoveEqLeft \text{\bf Subject to:}\sum_{j: \langle i, j \rangle \in \mathcal{E} }u_{ij}(t) \leq \rho X_i(t), \ \forall i \in \mathcal{V}; \\
\label{eqn:opt5_2}
\MoveEqLeft \qquad\qquad \ \sum_{i: \langle i, j \rangle \in \mathcal{E}  }u_{ij}(t) = (1-X_j(t))\Delta_j(t), \ \forall j \in \mathcal{V}; \\
\label{eqn:opt5_3}
\MoveEqLeft \qquad\qquad \ \sum_{i: \langle i, j \rangle \in \mathcal{E} } u_{ij}(t) C_{ij} \geq \delta_j(t) (1-X_j(t)), \ \forall j \in \mathcal{V}.
\end{alignat}
Constraint (\ref{eqn:opt5_2}) indicates that if  a household has no demand at time $t$, then it does not need incoming video traffic. For  the two-hop and non-splittable relay routing problem defined in (\ref{eqn:op2}) through (\ref{eqn:opt2_5}), we can change all design variables and demands to be time-dependent, and update  (\ref{eqn:opt2_2}) as:
\[
 \sum_{i: \langle i, j \rangle \in \mathcal{E}  } u_{ij}(t) = Y_j(t)+\Delta_j(t)(1-X_j(t)-Y_j(t)), \ \forall j \in \mathcal{V},
\]
which says that node $j$ needs to download video through exactly one incoming link if either $j$ is a relay node ($Y_j(t)=1$), or it is a terminal node $(X_j(t)=Y_j(t)=0)$ and has demand $(\Delta_j(t)=1)$. For the splittable relay routing problem defined in (\ref{eqn:op3}) through (\ref{eqn:op3-4}), it is sufficient to directly replace $\{f_{ij}, l_{si},\delta_i\}$ with time-dependent variables/constants $\{f_{ij}(t), l_{si}(t),\delta_i(t)\}$. Similar modifications can be made for formulations with LTE in Section \ref{subsec:optB}.
\subsubsection{Static Provisioning of Satellite Antennas and Dynamic Relay Routing}
\label{subsubsec:op8}
In this case, variables reflecting the positions of satellite antennas $\{X_i, l_{si}\}$ are time-independent, while the other variables are time-dependent, i.e., $\{Y_i(t), u_{ij}(t), f_{ij}(t), \lambda_{ij}(t)\}$. We can quickly convert the dynamic formulations in the previous section into the corresponding semi-dynamic formulation. For example, for the one-hop and non-splittable relay routing problem defined in (\ref{eqn:op5}) through (\ref{eqn:opt5_3}), we can have the semi-dynamic version as:
\begin{alignat*}{2}
\MoveEqLeft[2]\underset{\{X_i, u_{ij}(t)\}}{\text{\bf Minimize:}} \sum_{i \in \mathcal{V}} X_i \\
\MoveEqLeft \text{\bf Subject to:}\sum_{j: \langle i, j \rangle \in \mathcal{E} }u_{ij}(t) \leq \rho X_i, \ \forall i \in \mathcal{V}, t=1,\cdots, T; \\
\MoveEqLeft \qquad\qquad \ \sum_{i: \langle i, j \rangle \in \mathcal{E}  }u_{ij}(t) = (1-X_j)\Delta_j(t), \ \forall j \in \mathcal{V}, t=1,\cdots, T; \\
\MoveEqLeft \qquad\qquad \ \sum_{i: \langle i, j \rangle \in \mathcal{E} } u_{ij}(t) C_{ij} \geq \delta_j(t) (1-X_j), \ \forall j \in \mathcal{V}, t=1,\cdots, T.
\end{alignat*}
Similar modifications can be made for all other formulations in the Section \ref{subsec:optA} and \ref{subsec:optB}.

\subsubsection{Static Provisioning of Satellite Antennas and Static Relay Routiing}
\label{subsubsec:op9}
In this scenario, all design variables are time-independent, only the demand constants $\{\delta_i(t),\Delta_i(t)\}$ are time-dependent. All the formulations in the dynamic case can be modified accordingly. For example, the one-hop and non-splittable relay case become:
\begin{alignat*}{2}
\MoveEqLeft[2]\underset{\{X_i, u_{ij}\}}{\text{\bf Minimize:}} \sum_{i \in \mathcal{V}} X_i \\
\MoveEqLeft \text{\bf Subject to:}\sum_{j: \langle i, j \rangle \in \mathcal{E} }u_{ij} \leq \rho X_i, \ \forall i \in \mathcal{V}, t=1,\cdots, T; \\
\MoveEqLeft \qquad\qquad \ \sum_{i: \langle i, j \rangle \in \mathcal{E}  }u_{ij} = (1-X_j), \ \forall j \in \mathcal{V}, t=1,\cdots, T; \\
\MoveEqLeft \qquad\qquad \ \sum_{i: \langle i, j \rangle \in \mathcal{E} } u_{ij} C_{ij} \geq \delta_j(t) (1-X_j), \ \forall j \in \mathcal{V}, t=1,\cdots, T.
\end{alignat*}

\section{Approximation Algorithms}
\label{sec:Solution} 
In Section \ref{sec:optimization}, different scenarios are modeled either as binary programming or mixed-integer programming problems, which are both NP-hard problems. When the network size is small, one can use various optimization tools, such as CVX in MATLAB~\cite{rf24,rf25}, to get the exact optimal provisioning and relay routing solutions. However, when the network size is large, the computation time might become prohibitive. In this section, we develop heuristic approximation algorithms to obtain close-to-optimal solutions for large networks.  
%

The problem formulations in Sections \ref{subsubsec:op1} and \ref{subsubsec:op2} are similar to the classic set cover problem. Our objective is to determine the minimum number of nodes that can cover all other nodes in a given directed graph $\mathcal{G}$ with limited link capacity. Let $\mathcal{A}$ denote the relay matrix, where $\mathcal{A}[i,j]=1$ if and only if there is a wireless relay link from node $i$ to $j$, and the capacity of link $\langle i, j \rangle$ is larger than $\delta_j$, the total video demand of $j$.  Let $\mathcal{B}(i) \triangleq \{j \in \mathcal V: \mathcal{A}[i,j]=1\}$ be the set of nodes that can potentially download their TV demands from node $i$. Then call  $\mathcal{B}(i)$ the bin of node $i$. 

The One-hop and Non-splittable Relay problem formulated in Section~\ref{subsubsec:op1} can be approximately solved using the greedy heuristic  algorithm defined in Algorithm~\ref{alg:algo1}. Let $\mathcal{S}$ be the set of chosen source nodes, and $\mathcal{T}$ the set of terminal nodes that receive their TV channels from some source node in $\mathcal{S}$. At each iteration, node $i$ with the largest bin size is selected as a new source node. All nodes in node $i$'s bin are added to the terminal node set $\mathcal{T}$. If $i$'s bin has more than $\rho$ nodes, then we randomly select $\rho$ nodes to be covered by $i$. All the nodes in $i$'s bin are added to the terminal node set. All links from $i$ to its receivers are added to the relay topology. Our problem is different from the traditional set cover problem as each element of a bin has its own bin. Thus, after selecting a node as source, the nodes in its bin are not removed from the network, because they can still act as sources for other nodes in future iterations. As a result, when we select a new source, it might have been covered by some source node and added to the terminal set in previous iterations. We need to remove it from the terminal node set (line 10), and also remove its incoming video link from the relay topology (line 11). After we update the source and terminal node sets, all links going to source and terminal nodes no longer need to be considered, and thus are removed from the relay matrix. After the iterations, nodes those are not marked as either source or terminal node are isolated nodes that need satellite antennas. Finally, the relay topology and source set are returned.     
\begin{algorithm}
 \caption{Greedy algorithm for one-hop non-splittable relay}
 \label{alg:algo1}
 \begin{algorithmic}[1]
 \REQUIRE Relay matrix ($\mathcal{A}$) \\
 \ENSURE Satellite antennas positioning and one-hop relay topology \\
 \STATE \textbf{Initialization:} $\mathcal{S} \gets \phi$, $\mathcal{T} \gets \phi$, $\mathcal{A}_{tmp} \gets \mathcal{A}$, $\mathcal{A}_{opt} \gets \phi$ \\ 
 \WHILE {$\mathcal{A}_{tmp}$ is not empty}
 \STATE \textit{Calculate} the bin of each node based on $\mathcal{A}_{tmp}$, and find node $i$ with the largest bin. 
 \STATE $\mathcal{S} = \mathcal{S} \cup \{i\}$
 \IF {$|\mathcal B(i)| \le \rho$} 
 \STATE  $\mathcal{R}(i)=  \mathcal B(i)$
 \ELSE \STATE randomly select $\rho$ nodes in $\mathcal B(i)$ to $\mathcal{R}(i)$. 
 \ENDIF
 \STATE $\mathcal{T} = \mathcal{T} \cup \mathcal{R}(i) - \{i\}$
 \STATE  $\mathcal{A}_{opt} =  \mathcal{A}_{opt} \cup \{\langle i, k \rangle, \forall k \in \mathcal{R}(i)\} - \{\langle k, i \rangle, \forall k \in \mathcal{V}\} $
 \STATE $\mathcal{A}_{tmp} = \mathcal{A} - \{\mathcal{A}(m,n): m \in \mathcal{V}, n \in \mathcal{S} \cup \mathcal{T} \}$
 \ENDWHILE
 \RETURN relay topology $\mathcal{A}_{opt}$  and source set $\mathcal{S}_{opt} = (\mathcal{V} - \mathcal{S} - \mathcal{T}) \cup \mathcal{S}$
 \end{algorithmic}
\end{algorithm}
\setlength{\textfloatsep}{0pt}
 
Algorithm~\ref{alg:algo1} can be extended to cover the two-hop non-splittable relay case. Similar to the one-hop case, we develop a greedy iterative algorithm. At each iteration, we add node $i$ with the largest number of one-hop children as a new source. The links from node $i$ to their children $\mathcal{R}(i)$ are added to the relay topology. Different from the one-hop case, some nodes in $\mathcal{R}(i)$ might further act as relays and forward video to two-hop children of $i$. Let $\mathcal{D}(i,\mathcal{R}(i))$ be the set of nodes connecting to $i$ through $\mathcal{R}(i)$, i.e.,
\[\mathcal{D}(i, \mathcal{R}(i)) \triangleq \{k \in \mathcal V: \exists j \in \mathcal{R}(i) \text{ such that } C_{jk}  \ge \delta_k\}.\] 
Note, a node $k \in \mathcal{D}(i, \mathcal{R}(i))$ might connect to $i$ through multiple relay nodes in $\mathcal{R}(i))$, and it can be added as a two-hop child of $i$ through any one of them in the relay topology. To build the two-hop relay tree rooted at $i$, we develop another greedy iterative algorithm. Due to the space limit, we only give the algorithm sketch as follows. 
\begin{enumerate}
\item We first build the one-hop relay tree from $i$ to $\mathcal{R}(i)$, and update the spare capacity on link $\langle i, j \rangle, j \in \mathcal{R}(i)$ as $\tilde C_{ij} = C_{ij} - \delta_j$. 

\item We select the node, say $j_0$, with the highest spare capacity from node $i$ to grow the second hop relay. 

\item Among all children of $j_0$, we first select a node $k$ that is connected to $i$ only through $j_0$, if no such a node exists, we randomly select a child $k$ of $j_0$. If $\tilde C_{ij_0} \ge \delta_k$, we add $k$ as a two-hop child of $i$ through $j_0$ in the relay topology, and update the spare capacity $\tilde C_{ij_0} =  \tilde C_{ij_0} - \delta_k$. If no child of $j_0$ can be added to the relay topology, we set $\tilde C_{ij_0} =  0$.

\item Go back to Step 2, \emph{unless} either the spare capacity of all first-hop links originated from node $i$ become zero, or all nodes in  $\mathcal{D}(i, \mathcal{R}(i))$ are added to the relay topology. 
\end{enumerate}
After we build the two-hop relay tree rooted as node $i$, we move on to find the next source with the highest degree until all the nodes are covered. 
\vspace{-2mm}
\section{Performance Evaluation}
\label{sec:results}
We evaluate the proposed WiLiTV architecture using real household topology and user demand data from four communities with different household sparsity served by a major service provider in the USA. Community (i) consists of $22$ nodes with households sparsely located. Communities (ii), (iii) and (iv) consists of $21$, $13$ and $17$ households, respectively. The average distance between households in the community (i) is around 75m, while in the latter three communities it is less than 55m. The case with the LTE Base Station is considered for community (i) only, and its location and the available LTE resources are taken from the database of BS for that community.   
Using the optimization formulations in Section \ref{sec:optimization}, we find the optimal source provisioning and relay topologies under different relay routing complexity constraints. We further explore the use of parallel streams supported in IEEE 802.11n. Using beam-forming and Multiple Input, Multiple Output (MIMO) antenna techniques, up to four parallel streams can be supported in IEEE 802.11n. 

\begin{figure}[t]
\centering
\includegraphics[width=0.45\textwidth]{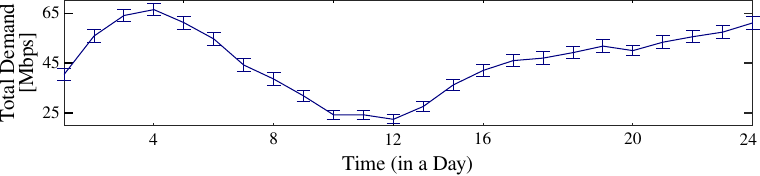}
\caption{Demand pattern in community (i): mean total demand per hour (50 samples) with vertical bars denoting 95\% confidence intervals.}
\label{fig:pic10}
\end{figure}

\begin{figure*}
         \centering
         \begin{subfigure}[b]{0.25\textwidth}
                 \includegraphics[width=\textwidth]{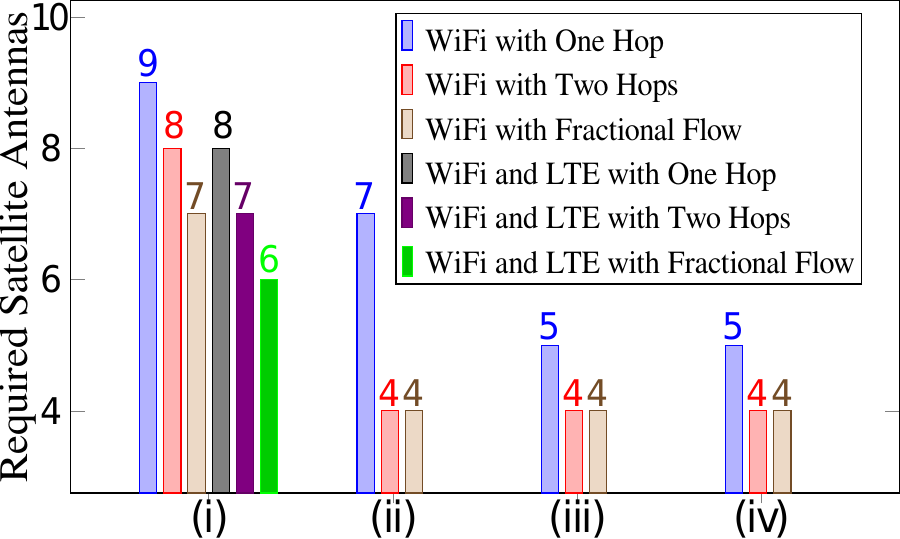}
                 \caption{Number of Streams = 1}
                 \label{fig:pic2}
         \end{subfigure}%
         \begin{subfigure}[b]{0.25\textwidth}
                 \includegraphics[width=\textwidth]{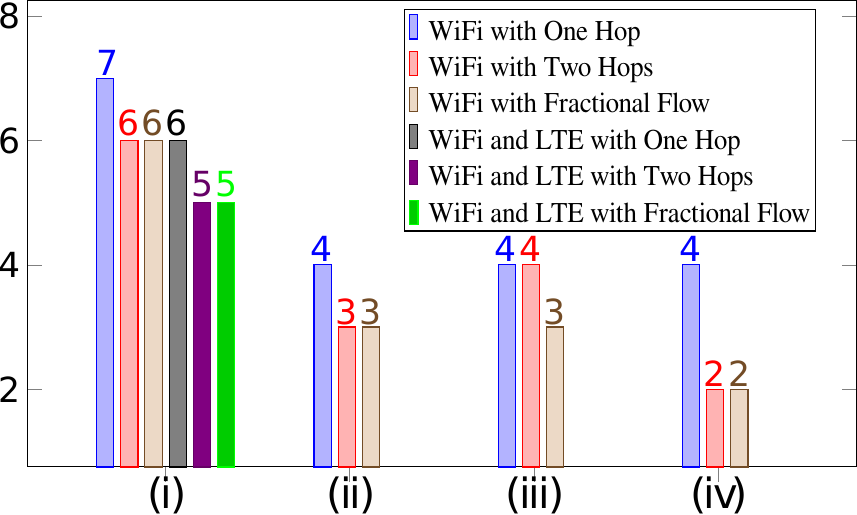}
                 \caption{Number of Streams = 2}
                 \label{fig:pic3}
         \end{subfigure}%
         \begin{subfigure}[b]{0.25\textwidth}
		 \includegraphics[width=\textwidth]{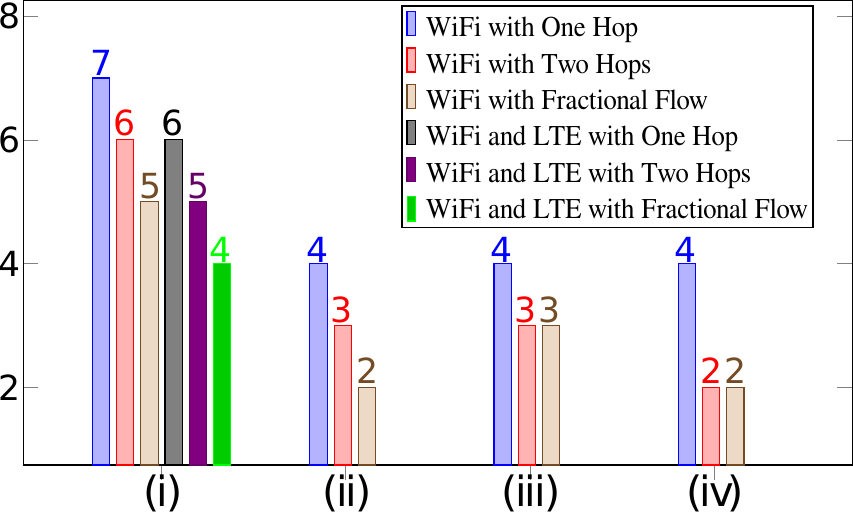}
                 \caption{Number of Streams = 3}
                 \label{fig:pic4}
         \end{subfigure}%
         \begin{subfigure}[b]{0.25\textwidth}
                 \includegraphics[width=\textwidth]{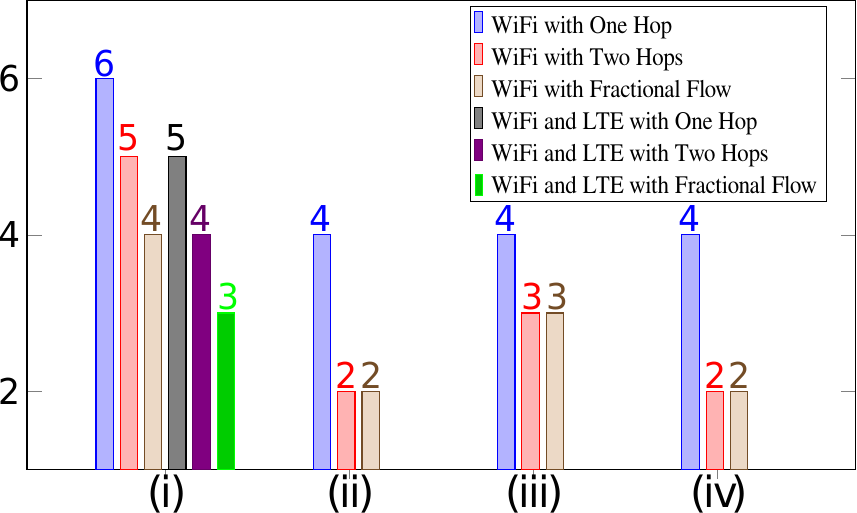}
                 \caption{Number of Streams = 4}
                 \label{fig:pic5}
         \end{subfigure}%
           \vspace{-2mm}
         \caption{Required number of satellite antennas in four considered communities to satisfy the maximum (fixed) user TV demands.}
  \label{fig:reqSat}
\end{figure*}

\begin{figure*}
         \centering
         \vspace{-3mm}
         \begin{subfigure}[b]{0.25\textwidth}
                 \includegraphics[width=\textwidth]{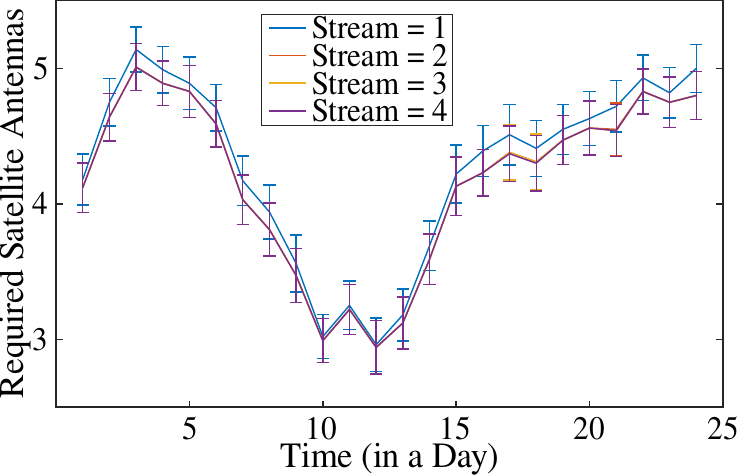}
                 \caption{WiFi over one hop}
                 \label{fig:pic6}
         \end{subfigure}%
         \begin{subfigure}[b]{0.25\textwidth}
                 \includegraphics[width=\textwidth]{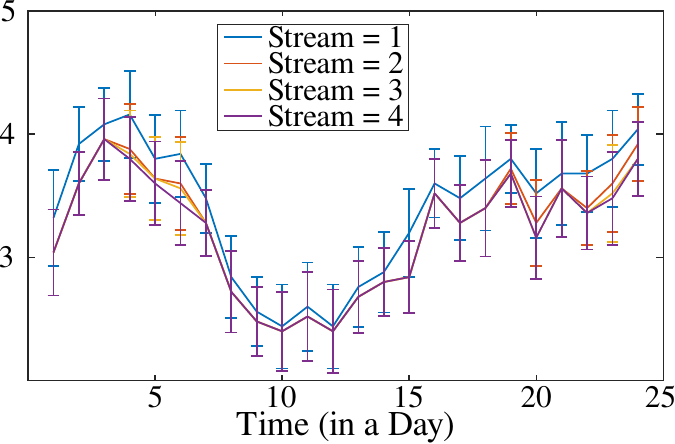}
                 \caption{WiFi over two hops}
                 \label{fig:pic7}
         \end{subfigure}%
         \begin{subfigure}[b]{0.25\textwidth}
		 \includegraphics[width=\textwidth]{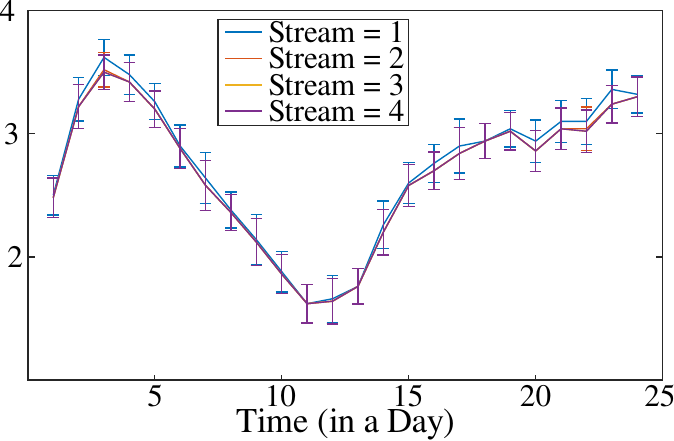}
                 \caption{WiFi and LTE over one hop}
                 \label{fig:pic8}
         \end{subfigure}%
         \begin{subfigure}[b]{0.25\textwidth}
                 \includegraphics[width=\textwidth]{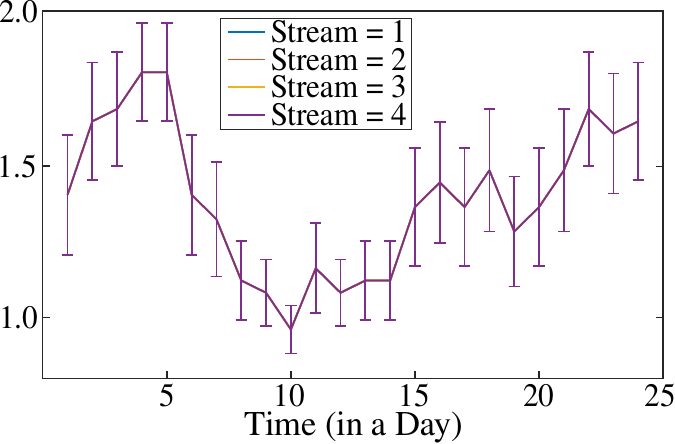}
                 \caption{WLAN and LTE over two hops}
                 \label{fig:pic9}
         \end{subfigure}%
         \vspace{-2mm}
         \caption{Community (i): variation of required satellite antennas with non-splittable relay to satisfy demand at each time instant.}
  \label{fig:reqvar}
\end{figure*}
\begin{figure*}
         \centering
          \vspace{-3mm}
         \begin{subfigure}[b]{0.25\textwidth}
                 \includegraphics[width=\textwidth]{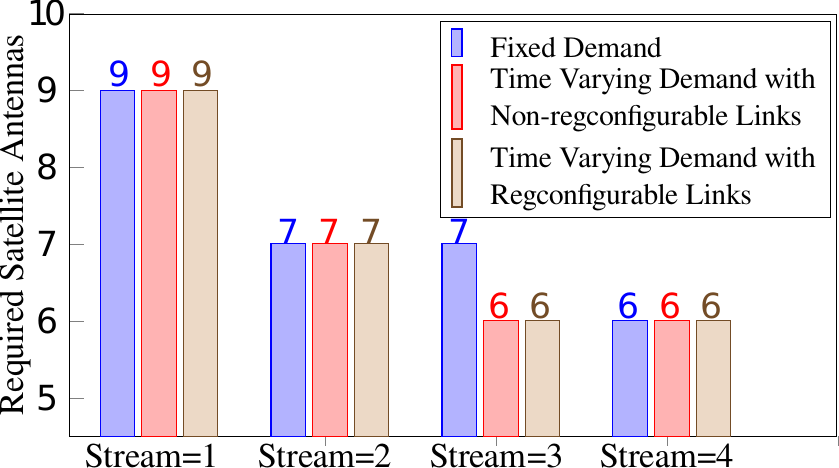}
                 \caption{WiFi over one hop}
                 \label{fig:pic11}
         \end{subfigure}%
         \begin{subfigure}[b]{0.25\textwidth}
                 \includegraphics[width=\textwidth]{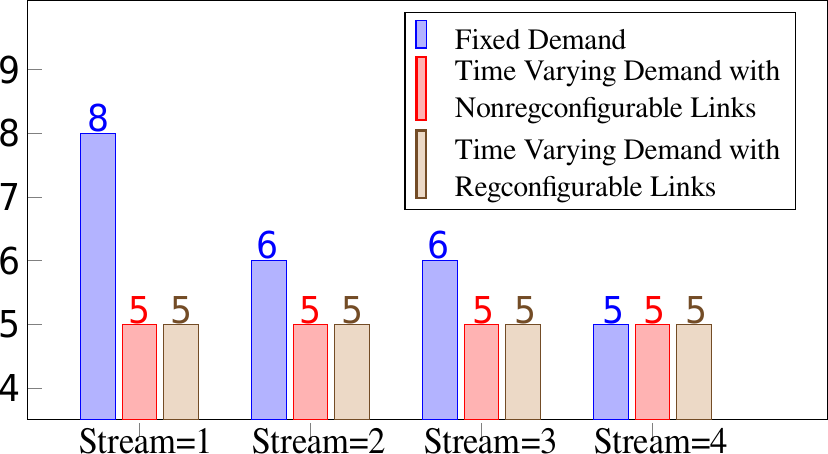}
                 \caption{WiFi over two hops}
                 \label{fig:pic12}
         \end{subfigure}%
         \begin{subfigure}[b]{0.25\textwidth}
		 \includegraphics[width=\textwidth]{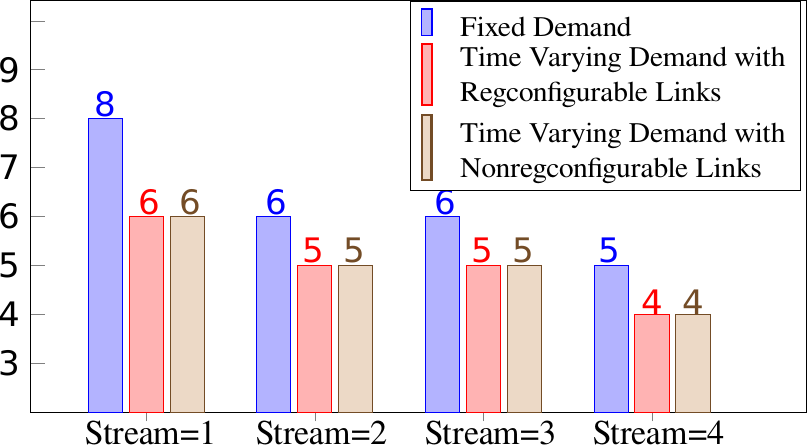}
                 \caption{WiFi and LTE over one hop}
                 \label{fig:pic13}
         \end{subfigure}%
         \begin{subfigure}[b]{0.25\textwidth}
                 \includegraphics[width=\textwidth]{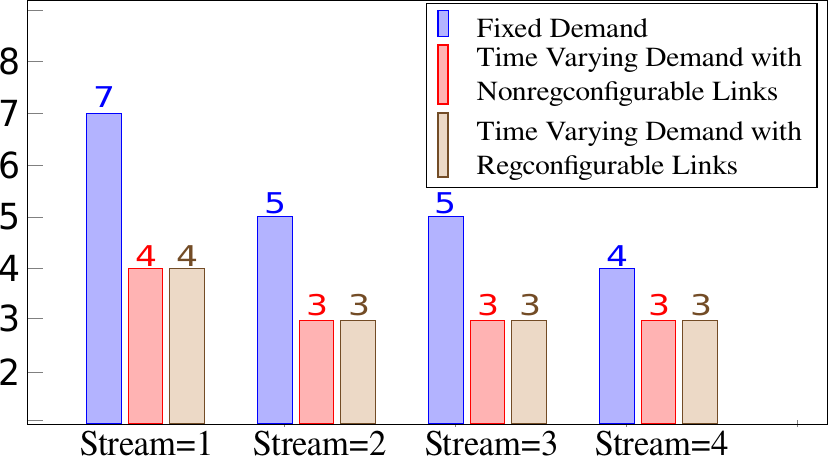}
                 \caption{WiFi and LTE over two hops}
                 \label{fig:pic14}
         \end{subfigure}%
         \vspace{-2mm}
         \caption{Community (i): required number of satellite antennas with non-splittable relay to satisfy demands at all time instants.}
  \label{fig:Satreq}
  \vspace{-5mm}
\end{figure*}

Fig.~\ref{fig:pic10} shows a typical demand pattern over a day. From this real demand data, obtained from a service provider, we compute the probability distributions for the requested number of TV channels for each household at each hour of a day. In each simulation, we draw user demand samples from these probability distributions. The carrier frequencies for WiFi and LTE are $5$ GHz and $2$ GHz respectively, the channel bandwidth for both WiFi and LTE are $20$ MHz. The degree of connectivity of household is $\rho=5$. The data rate of each TV channel is $5$ Mbps, and the maximum relay hop count is $h=2$. For these simulation parameters, our heuristic algorithm obtains the identical result with a shorter computation time (particularly in the two-hop scenario, where the exact algorithm takes tens of seconds while our heuristic algorithm takes a few seconds).  

\vspace{-2mm}
\subsection{Fixed Demand}
\label{subsec:resultfix}
First, we consider \emph{source provisioning} and \emph{relay routing} using peak demand per household observed over a long period of time. The fixed peak demand scenario gives us an upper bound on the required number of satellite antennas. In Fig.~\ref{fig:reqSat}, we present the number of nodes (households or LTE BSs) that must be equipped with satellite antennas for live TV content distribution in different scenarios for different communities. 
With one IEEE 802.11n stream and one-hop relay, 50\%, 30\%, 38\% and 29\% nodes must be equipped with satellite antennas in the four communities, respectively. Similarly, with four parallel IEEE 802.11n streams and one-hop communication, 36\%, 19\%, 30\% and 23\% nodes must be equipped with satellite antennas. When LTE BSs are available, the required number of satellite antennas further decreases. In the best case scenario, with a heterogeneous network consisting of both LTE and WiFi links over a two-hop relay with fractional flow, the required number of satellite antennas are 13\% for community (i) for four streams. Similarly for four streams using WiFi links over two hops, the required number of satellite antennas reduces to 9\%, 23\% and 11\% for community (ii), (iii) and (iv) respectively. {\it This suggests that additional WiFi link capacities resulting from more streams directly translate  into cost savings on satellite antennas, especially with two-hop relays} Also notice that, even though fractional flows are more flexible than all-or-nothing flows, in the evaluated scenarios, they bring no or marginal performance gains over the corresponding all-or-nothing flows. {\it This suggests that non-splittable relay routing may achieve most of the cost savings in practice, without incurring the complexity and reliability concerns of splittable relay routing.}
\vspace{-2mm}
\subsection{Dynamic Solution for Time-varying Demand}
\label{subsec:timeVarying}
Fig. \ref{fig:reqvar} plots the required number of satellite antennas for time-varying demand with impractical dynamic antenna provisioning and dynamic relay routing as studied in Section~\ref{subsubsec:op7}. We consider non-splittable relay routing  to determine the variation of the required satellite antennas. They serve as the lower bounds for satisfying time-varying demands. The vertical bars in the figure denote the 95\% confidence intervals. We can observe that the result highly depends upon the TV content demand. Similar to previous observation for fixed demand, {\it the required numbers of satellite antennas decrease from one-hop relay to two-hop relay}. One can also observe that the confidence intervals with two-hop relays are  significantly higher than one-hop relay. This suggests that {\it the additional gain from two-hop relays are more time-dependent.} For two-hop WiFi, we can observe that there is a larger gap between one stream and two streams. This observation can simply be explained from the fact that {\it a higher number of streams is useful to support multi-hop communications}. Similar results are obtained for fractional flows with both WiFi and heterogeneous networks. 
\vspace{-2mm}
\subsection{Static Satellite Antenna Provisioning}
Finally, an optimal distribution topology is obtained with static satellite antennas having, (i) reconfigurable and, (ii) non-reconfigurable links, as discussed in Sections~\ref{subsubsec:op8} and \ref{subsubsec:op9}. Fig.~\ref{fig:Satreq} compares the required numbers of satellite antennas to satisfy time varying demands at all time instants to results obtained for fixed peak TV demands in Fig.~\ref{fig:reqSat}, all with non-splittable relays. {\it Fig.~\ref{fig:reqSat} suggests that formulations considering the time-varying demands, instead of per-user peak demands, can lead to a higher cost saving in two-hop and LTE cases}. For the two cases considered with fixed positioning of satellite antennas with configurable and non-reconfigurable relays, we obtained the same results for all the combinations studied. {\it This suggests that fixed positioning of satellite antennas and proper selections of relay routing are sufficient for an optimal distribution topology}.  
\vspace{-2mm}
\section{Conclusion}
\label{sec:conclusion}
We provide an all wireless solution to deliver live TV services. Some service providers now have the benefit of leveraging multiple access technologies to distribute live TV content (e.g. satellite, WiFi and LTE). We capitalize on this opportunity to create a distribution infrastructure that is optimized to serve large residential neighborhoods with a minimal number of TV content injection points. We solve multicommodity optimization flow problems to model various scenarios. Using real data from a national TV service provider, we show that our proposed architecture can save provisioning costs by between 75\% to 90\%. We experiment using four different representative residential neighborhoods with time-varying traffic demands. Our investigation shows that there is an optimum strategy for placing the satellite dish antennas combined with an appropriate selection of WiFi routes to meet the time-varying demand of subscribers/households.
\vspace{-2mm}
\bibliographystyle{IEEEtran}
\bibliography{references}
\end{document}